\documentclass{ws-p9-75x6-50}

\newcommand{\grad}{\nabla}
\newcommand{\fpi}{f_\pi}
\newcommand{\kf}{k_{\scriptscriptstyle\rm F}}
\newcommand{\nab}{\overrightarrow{\nabla}}
\newcommand{\galnab}{\stackrel{\leftrightarrow}{\nabla}}

\begin{document}

\title{Recent Developments in the\\ Nuclear Many-Body Problem}

\author{R. J. Furnstahl}

\address{Dept.\ of Physics, Ohio State University, Columbus, OH 43210, 
         USA\\E-mail: furnstahl.1@osu.edu}

%\author{More Authors}

%\address{Another address}


\maketitle

\abstracts{%
The study of quantum chromodynamics (QCD) over the past quarter 
century has had 
relatively little impact on the traditional approach to
the low-energy nuclear many-body problem.
Recent developments are changing this situation.
New experimental capabilities and theoretical approaches are opening
windows into the richness of many-body phenomena in QCD.
A common theme is the use of effective field theory (EFT) methods,
which exploit the separation of scales in physical systems.
At low energies,
effective field theory can explain how existing 
phenomenology emerges from QCD and how to refine it systematically.
More generally, the application of EFT methods to  many-body
problems promises insight into the analytic structure of observables,
the identification of new expansion parameters, and
a consistent organization of many-body corrections,
with reliable error estimates.
}

\section{Introduction}
At a fundamental level, atomic nuclei are described by quantum
chromodynamics (QCD) with colored quark and
gluon degrees of freedom, 
whose interactions are asymptotically free at short distances.
Yet under ordinary conditions, colorless nucleons in a nucleus largely
retain their identity \cite{CROSSING}.   
In the traditional approach to the low-energy nuclear many-body problem,
a phenomenological two-nucleon potential is fit to scattering data 
and properties of the deuteron, and 
solutions to the
many-body Schr\"odinger equation for nuclei across the periodic table and
nuclear matter are approximated by various sophisticated
methods.
Three-body forces and meson-exchange currents
are added only when required by discrepancies with data.
The study and validation of QCD in other contexts
over the past quarter century has had 
relatively little impact on this phenomenology \cite{CROSSING}.

Recent developments are changing this situation in two major ways.
First,
the nuclear many-body problem has become the QCD many-body problem,
involving explorations of phenomena throughout the phase diagram of QCD.
Second, effective field theory (EFT) methods are being used
to build 
bridges from QCD to traditional nuclear many-body phenomenology.
The phrase ``effective theory'' has often  denoted a model
used because one couldn't solve the underlying theory.  
In contrast, an EFT
is a field theory that reproduces the results of an 
underlying theory
in a systematic and model-independent way, but in a limited domain.

It would be impractical to make a complete survey of this broad range
of activity, so 
we will instead present a coarse-grained tour with selected stops.
Many more details are available in the cited references.
We will start with ``teasers'' 
from two of the many new frontiers in exploring the QCD
phase diagram, many-body physics at small $x$ and color superconductivity.
New insights into established approaches within the traditional framework
and how successful phenomenology emerges from low-energy QCD are considered
next.
Finally, we give an example of how new ideas for attacking many-body
problems arise from applying the EFT perspective.

\section{Exploring the QCD Phase Diagram}

Quantum chromodynamics (QCD) is a gauge theory of SU(3) color
charges, with quarks and gluons as the fundamental degrees of freedom.
It has many analogies to quantum electrodynamics (QED), but important
differences follow from the non-abelian structure.  In particular, 
gluons carry the color charge and interact with each other.  Two
prominent consequences are asymptotic freedom (the interaction becomes
weak at high energies/short distances) and color confinement
(observed hadrons have no net color charge).
Quarks come in six varieties, called flavors, with masses that are 
very small compared to the relevant QCD scale (up, down) 
or very large (charm, top, bottom).
The strange quark mass is somewhere in the middle; 
whether it should be considered heavy or light has an
important influence on the phase structure of QCD.

\begin{figure}[t]
%\figurebox{22pc}{15pc}{} % to have a box alone
\caption{The conjectured phase diagram of QCD
(see Ref.~\citen{HEINZ} for details).  \label{fig:phasediag}}
\epsfxsize=3.0in % will enlarge or reduce the postscript figures based on the xsize
\centerline{\epsfbox{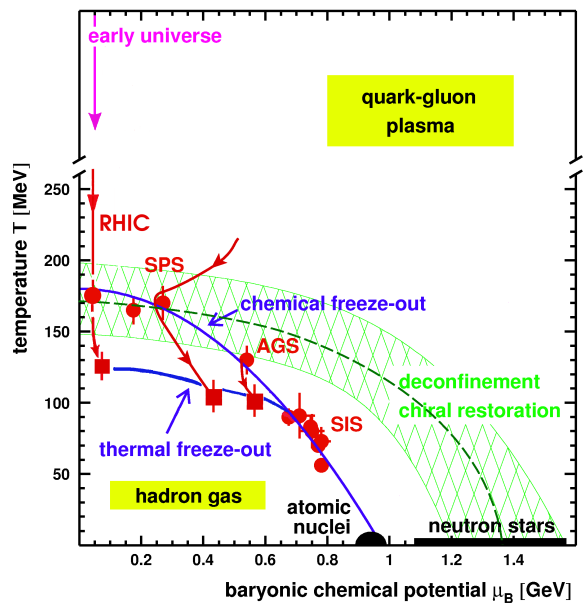}} % postscript image file name
\end{figure}

Figure~\ref{fig:phasediag} is  a conjectured phase diagram for QCD.
The axes here are the temperature and the chemical potential for baryon
number.  The traditional nuclear many-body problem lives in the small
black blob at low temperature with chemical potential near one GeV.
But now the nuclear many-body problem encompasses the entire plane
and can more correctly be termed the QCD many-body problem.  Even the
physical vacuum ($T=0$, $\mu_B=0$) 
is an extraordinarily complex, coherent many-body
medium characterized by nonperturbative quark and gluon 
condensates \cite{RHIC}.

One of the main tasks of the new nuclear many-body problem is to explore
the phase structure in Fig.~\ref{fig:phasediag}, 
such as the expected deconfinement and chiral symmetry restoration
phase transitions.  
These transitions are predicted by lattice simulations; however, the 
simulations are
restricted at present to zero chemical potential.
Energetic collisions of heavy nuclei are being used to
probe different regions of temperature and chemical potential; the figure
shows data points from several heavy-ion experiments.  The latest
accelerator, RHIC, is now running and producing data \cite{RHIC,HEINZ}.

There are many interesting challenges for many-body theory implicit
in Fig.~\ref{fig:phasediag}.  Here we will
briefly consider only two: small-$x$ physics and color superconductivity.

\subsection{Many-Body Physics at Small $x$}\label{subsec:smallx}

The initial conditions of
ultra-relativistic heavy-ion collisions strongly influence 
the details of quark-gluon plasma formation.
Particularly important are 
the ``wee'' parton distributions in nuclei, which are virtual excitations of the
nuclear ground state (mostly gluons) carrying a  
small fraction $x$ of the longitudinal momentum of each colliding 
nucleus \cite{VENUGOPALAN}.
Large energy $s$ and momentum transfer $Q^2$ but small $x \propto Q^2/s$ is the
regime of high virtual gluon densities.
Perturbative QCD evolution predicts rapid growth in the gluon distribution as
$x\rightarrow 0$, but eventually gluon recombination and screening effects
should become important.  
An important question is whether the gluon density saturates
and at what scale \cite{VENUGOPALAN}.

The large occupation number of gluon states implies that a classical effective
field theory (EFT) is a good starting point, with corrections from
a nonlinear renormalization group analysis.
The characteristic energy scale at RHIC and planned future accelerators
is high enough that the effective coupling is weak, although the
physics is nonperturbative. 
McLerran and collaborators have predicted Bose-Einstein condensation at
sufficiently high gluon density, leading to a new state of matter they
call a ``colored glass condensate.''
With respect to natural time scales,
the color fields evolve slowly and are disordered\cite{MCCLARREN,IANCU}.  
The EFT for this system, 
formulated in the infinite momentum frame (in light cone gauge)
is a (nearly) two-dimensional theory with a structure mathematically analogous
to that of a disordered system of Ising spins in a random magnetic 
field \cite{MCCLARREN}.
Over the coming decade, ongoing work
on the theory of high density QCD will elucidate experimental signatures of this
condensate \cite{RHIC}.

\subsection{Color Superconductivity}
Could the interior of a neutron star be a quark-gluon fluid,
as implied by Fig.~\ref{fig:phasediag}?
It would seem so based on a naive argument:  asymptotic freedom implies that the
force between quarks weakens as the momentum scale of interaction increases,
and
the behavior at low temperature and high density is determined by high-momentum
quarks at the Fermi surface, so we would expect matter to be a Fermi sea of
essentially free quarks.  However, this argument is well known to be too naive:
if there is quark-quark attraction, then a Cooper pairing instability
is inevitable.
In fact, an analysis using one-gluon-exchange (applicable at very high
density) or instanton-induced interactions (nonperturbative effects at
lower density) indicate attraction between quarks in the color anti-symmetric
$\overline 3$ state \cite{BAILIN,COLORSC,ALFORD}.

The methods of superconductivity have been readily adapted to this problem,
which has drawn particular interest because this high density
regime is theoretically tractable.
There is a rich and structured theory revealed
as one varies the number of colors $N_C$ and the number of quark flavors
$N_f$ (and their masses).
To assess the different possible instabilities, theorists have studied
the renormalization-group evolution of four-fermion operators using
the methods pioneered by Shankar and Polchinski \cite{SHANKAR}, 
in which one integrates
out states toward the Fermi surface.
The central result of the analysis is the identification of condensates
in diquark channels, analogous to Cooper pairs of electrons; this is called
color superconductivity \cite{COLORSC}.    

\begin{figure}[t]
%\figurebox{22pc}{15pc}{} % to have a box alone
\caption{Phase diagram for two  massless flavors \cite{ALFORD}.  
   \label{fig:colorsupertwo}}
\epsfxsize=3.48in % will enlarge or reduce the postscript figures based on the xsize
\centerline{\epsfbox{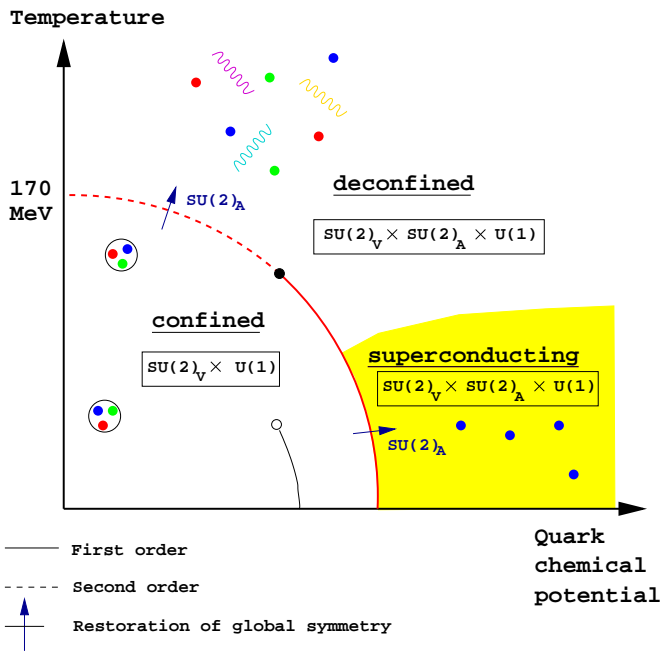}}
\end{figure}

\begin{figure}[t]
\caption{Phase diagram for three massless flavors \cite{ALFORD}.
  \label{fig:colorsuperthree}}
\epsfxsize=3.48in\centerline{\epsfbox{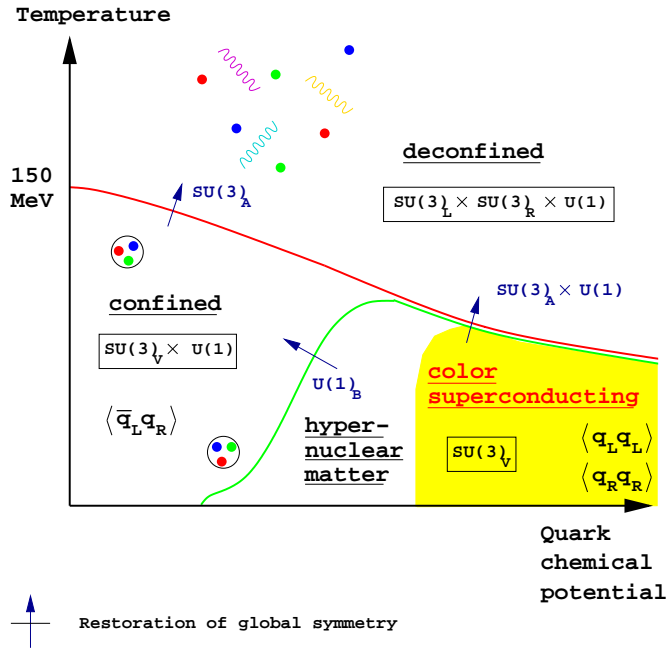}} 
\vspace*{-5pt}
    % postscript image file name
\end{figure}

To determine how the instability is resolved, a model energy functional
for the ground state is needed.  The simplest analyses use four-Fermi
interactions in a self-consistent mean-field (NJL) gap equation.
In the vacuum, a non-trivial solution gives a constituent mass for
quarks from spontaneous chiral symmetry breaking; calibrating this to
be about
400 MeV determines the coupling.  Then one studies the quantitative
phase structure \cite{ALFORD}.

For two massless flavors (so $m_s \rightarrow \infty$), the color
$\overline 3$, flavor singlet channel wins, and $SU(3)_{\rm color}$
is broken down to $SU(2)_{\rm color}$. 
The phase and symmetry structure is illustrated in 
Fig.~\ref{fig:colorsupertwo}.
The low-lying spectrum has five massive gluons and two colored massive
quarks with gap $\Delta$. 
There is a tri-critical point that survives the small up/down quark
masses, which may be observable in heavy-ion collisions as the analog
of critical opalescence \cite{COLORSC}.

For three massless flavors, the preferred mode is ``color flavor locking,''
which implies color and flavor are each broken, but combinations of
generators remain good symmetries.  There are eight massive gluons and all quark
modes are gapped.  
The phase and symmetry structure is illustrated in 
Fig.~\ref{fig:colorsuperthree}.
A fascinating possibility is quark-hadron continuity:
that these excitations map smoothly onto the hadronic phase.  For example,
the gluons become the octet
of vector bosons in hypernuclear matter.
A current challenge is to determine the correct interpolation of these
phases for real-world quark masses.

All analyses have revealed truly high-$T_c$ superconductors, with gaps of
order 100\,MeV (with current uncertainties plus or minus a factor of two)
and critical temperatures of order 50\,MeV or $10^{12}\,$K.
Since the attractive force comes from a primary interaction, color
superconductivity is particularly robust.  
The temperature in a neutron
star after a few seconds following its birth is much less than the
critical temperature, so if the interior density is high enough to support quark
matter, it will be a color superconductor.

Detecting the consequences of color superconductivity 
for neutron star physics is
a challenge, however.  The impact on the equation of state is ${\cal
O}(\Delta/\mu)^2$, which at a few percent is much less than the uncertainties in
the equation of state.  Thus the main signatures will be found elsewhere. 
Since
color superconductivity gives mass to low-lying excitations, one should look
toward transport phenomena (mean-free paths, conductivities, viscosities, and
the like).  One proposal is that the cooling rate of neutron stars, which is
determined by the heat capacity and emissivity, should show a characteristic
slow-down signature for the 2SC phase.  The color-flavor locking phase
might be detected by the time distribution of the neutrino pulse from
a supernova (such as 1987A).  
Future work will concentrate on sharpening these signals
and exploring other possibilities \cite{COLORSC}.

\section{From QCD to Nuclear Phenomenology}

How does traditional nuclear phenomenology relate to QCD?
Proposed paths from low-energy QCD to nuclei have included
models with explicit constituent quark degrees of freedom (e.g., a
quark shell model) and many-body skyrmions, in which nucleons appear
as solitons in the pion field.  However, these approaches
have had little quantitative
impact on our understanding of nuclear many-body physics and little
qualitative relation to the traditional approach.
In recent years a much more promising candidate has emerged: chiral effective
field theory.

\subsection{Chiral Effective Field Theory}

The effective field theory approach is grounded in some very general
physical principles \cite{LEPAGE}.
If a system is probed or interacts at low energies,
resolution is also low, and fine details of what happens at short
distances or in high-energy intermediate states are not resolved.
Therefore, the short-distance structure can be replaced by something
simpler without distorting the low-energy observables.
This is analogous to a multipole expansion, in which a complicated,
distributed charge or current distribution is replaced 
for long-wavelength probes by a series
of point multipoles.
The use of a local lagrangian provides a framework for carrying out
this program in a complete and systematic way.
The uncertainty principle implies that high-energy
intermediate states are highly virtual and only last for a short time,
so their effects are not distinguishable from those of local 
operators \cite{LEPAGE}.

The effective degrees of freedom (dof's) depend on a separation or resolution
momentum scale $\Lambda$.  Long-range dof's must be treated explicitly
while short-range physics is encoded in the coefficients of the local
operators.  For low-energy QCD at the momentum scales relevant for
bound nucleons, the appropriate degrees of freedom
(neglecting strangeness) are pions and
nucleons,
and $\Lambda \approx 500$--$1000\,$MeV.
The long-range pion physics is constrained by chiral symmetry, so
we have a chiral EFT, which has much in common with traditional
potential models \cite{CROSSING}.

There are three general ingredients of an EFT approach, which we will illustrate
with the particular details of the chiral EFT.  
Further explanations and other examples can be found in
Ref.~\citen{CROSSING}.
\begin{enumerate}
  \item {\em Construct the most general lagrangian density with appropriate
  low-energy degrees of freedom that is consistent with the global and
  local symmetries of the underlying theory.\/}
  For chiral EFT, this means 
  ${\cal L}_{\rm eft} = {\cal L}_{\pi\pi}
             + {\cal L}_{\pi N} +  {\cal L}_{NN}$, so nuclear
   many-body physics is united with pion-pion and pion-nucleon
   scattering physics.  Chiral symmetry allows us to treat long-distance
   pion physics systematically.
   \item {\em The declaration of a regularization and renormalization scheme.\/}
   Because the long-distance physics is insensitive to {\em details\/}
   of short-distance physics, the results are ultimately independent
   of the details of the scheme, but different approaches may have different
   convergence properties.  Also, the most useful EFT 
   formulation will depend on the few- or many-body system under 
   consideration.
   The most advanced chiral EFT
   calculations at present use a smooth momentum cutoff \cite{EPELBAUM}.
   \item {\em A well-defined power counting, which provides small expansion
   parameters.\/}  Power counting is a procedure to determine what graphs
   to include (or sum) at each order.  
   It ensures consistency with physical principles
   and conservation laws while allowing estimates of truncation errors.
   The separation of physical scales provides expansion
   parameters; for chiral EFT it is the momentum ${\bf p}$ and/or pion mass
   $m_\pi$ (generically called $Q$) divided by a chiral symmetry breaking
   scale $\Lambda_\chi \approx 1\,$GeV.  Weinberg's prescription is to power count
   the effective
   potential $V_{NN}$ and then solve the Schr\"odinger equation.
   Chiral symmetry implies 
   $V_{NN} = \sum_{\nu=\nu_{\rm min}}^{\infty} c_\nu Q^\nu$, where $\nu$
   is given by a formula based on the topology of a given diagram and
   the nature of its vertices.  The key 
   to a systematic expansion is a lower bound
   $\nu_{\rm min}$ \cite{VANKOLCK}.   
\end{enumerate}

\begin{figure}[t]
%\figurebox{22pc}{15pc}{} % to have a box alone
\caption{NN phase shift predictions in LO (dots), NLO (dashes), and NNLO
     (solid) 
     compared to experiment (diamonds).  The $x$--axis is the energy in GeV
     and the $y$ axis is in degrees.  \label{fig:phases}}
\epsfxsize=2.45in % will enlarge or reduce the postscript figures based on the xsize
\epsfclipon
\centerline{%
\epsfbox{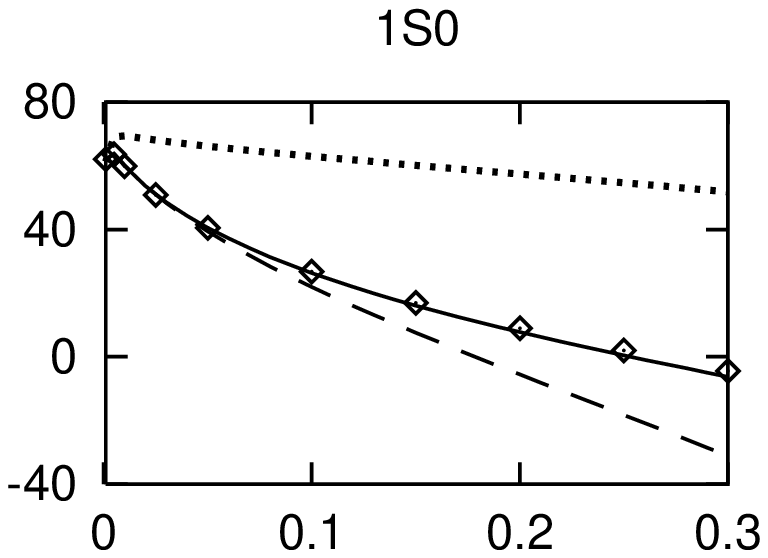}
\epsfxsize=2.45in % will enlarge or reduce the postscript figures based on the xsize
\epsfbox{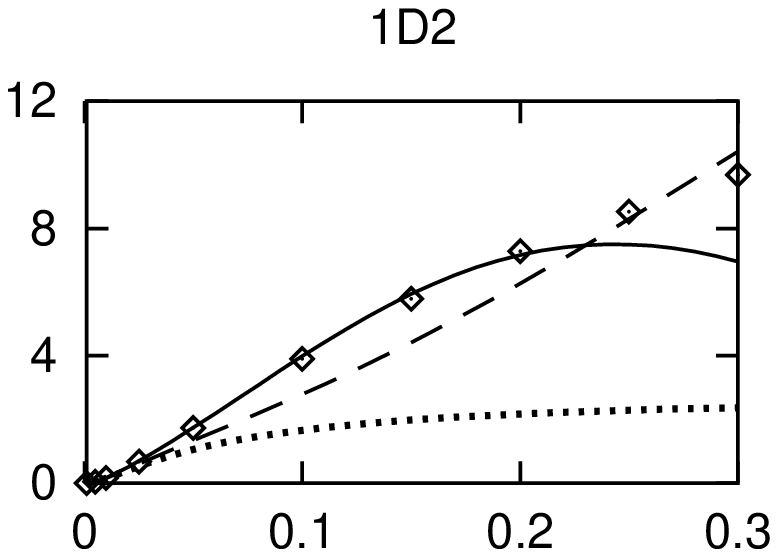} % postscript image file name
}
\end{figure}

The most complete application of this framework has been by
Epelbaum et al., who have adapted a unitary transformation method
to obtain an energy-independent potential \cite{EPELBAUM}.
Their recent results reveal an orderly progression in predictions
from
LO to NLO to NNLO (next-to-next-to-leading order), as shown for
two (of many) neutron-proton scattering phase shifts 
in Fig.~\ref{fig:phases} \cite{EPELBAUM}.
At each order, the power counting scheme specifies the new diagrams
needed; coefficients are determined from fits to $\pi N$ or $NN$
scattering data at low energy. 
The calculations 
become closer to experiment (better at low energy) and less sensitive
to the value of the momentum cutoff (which is 500--600\,MeV for NNLO),
in accord with the expectations of an EFT.

\begin{figure}[t]
\caption{Comparing coefficients from phenomenological models
  to chiral EFT \cite{EPELBAUM}.  
  The leftmost bar for each coefficient is LO (the length reflects the variation
  with cutoff), the middle bar is NLO, and the
  symbols are values extracted from the indicated boson-exchange potentials.  
  \label{fig:ressat}}
\epsfxsize=3.5in % will enlarge or reduce the postscript figures based on the xsize
\epsfclipon
\centerline{\epsfbox{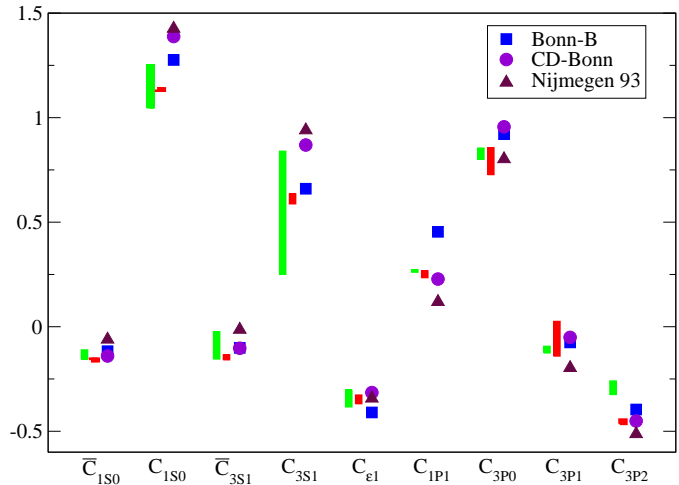}} % postscript image file name
\end{figure}

The coefficients in the lagrangian
can be estimated by naive dimensional analysis
(NDA), as originally proposed for low-energy QCD by Georgi and Manohar. 
Any term in the lagrangian with nucleon fields $N$ and pion fields
$\pi$ can be put in the form:
\begin{equation}
   {\cal L}_{\rm eft}
     \sim c_{lmn} \left( \frac{N^\dagger(\cdots)N}{\fpi^2 \Lambda_\chi}
               \right)^l
               \left(\frac{\pi}{\fpi}\right)^m
               \left(\frac{\partial^\mu,m_\pi}{\Lambda_\chi}\right)^n
               \fpi^2\Lambda_\chi^2 \ ,  \label{eq:NDA}
\end{equation}
where the pion decay constant $f_\pi \approx 100\,$MeV and the
chiral symmetry breaking scale $\Lambda_\chi \approx 1000\,$MeV.
If these scales are correctly incorporated, the remaining dimensionless
couplings $c_{lmn}$ should be of order unity; if so they are called
``natural''.
The explicit fits verify that the EFT is natural except for one combination
of constants that is unnaturally {\em small\/}.
This is a signature of a symmetry not explicitly considered; in this
case it is the Wigner $SU(4)$ spin-isospin symmetry \cite{EPELBAUM}.

The constants also reveal that
the chiral EFT potential is closely related to conventional NN potentials
based on boson exchange.  The latter can be considered to be models of the
short-distance physics that is unresolved in the chiral EFT (except for pion
exchange).  Therefore, this physics 
should be encoded in the coefficients of contact terms
in the EFT.  We can test this relation by comparing the chiral EFT
coefficients to corresponding coefficients derived from several boson-exchange
models in Fig.~\ref{fig:ressat}.  The pattern from the chiral EFT is a blueprint
of low-energy QCD and it is quantitatively reproduced by 
the phenomenology \cite{EPELBAUM}.
This is a major step
towards the construction of a path from QCD to the inputs of traditional 
nuclear many-body
theory!

\vspace*{.1in}

\begin{figure}[t]
%\figurebox{22pc}{15pc}{} % to have a box alone
\caption{Table of nuclides and corresponding theoretical approaches
  for different regimes of nucleon number $A$ \cite{RIAPAPER}.  
  \label{fig:nuclides}}
\epsfxsize=3.5in % will enlarge or reduce the postscript figures based on the xsize
\epsfclipon
\centerline{\epsfbox{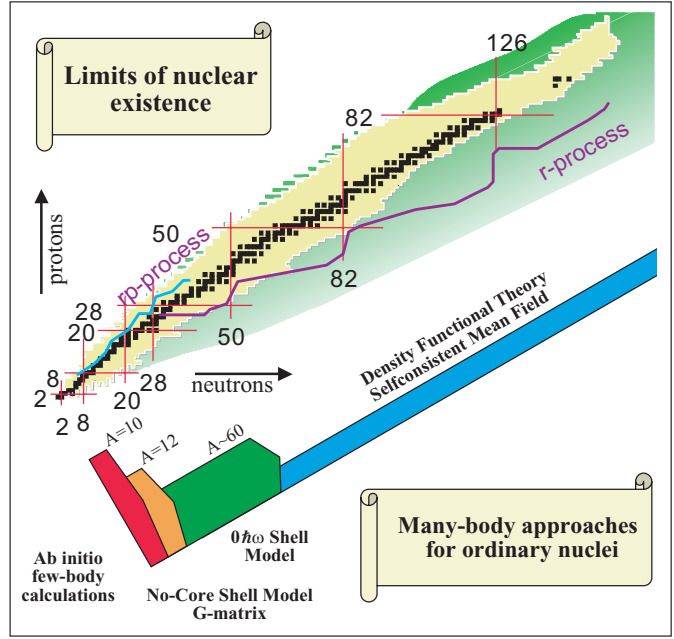}} % postscript image file name
\vspace*{-2pt}
\end{figure}

The theoretical many-body approaches for nuclei under ordinary conditions
are shown in Fig.~\ref{fig:nuclides}.  
For low $A$ (total number of protons plus neutrons), Monte Carlo methods
are able to calculate directly given free-space potentials. 
In the intermediate region shell-model methods are used, while in the
region of large $A$, where most of the nuclides lie, 
self-consistent mean-field models and related approaches
are most often applied.
We  consider the impact of EFT approaches on each in turn.

\subsection{Monte Carlo calculations of Light Nuclei}

Nuclei present a spin-isospin nightmare that make the cost of direct
Monte Carlo calculations
grow exponentially with $A$.
Through heroic efforts,
Carlson et al.\ have recently pushed
Green's function Monte Carlo  (GFMC) methods through $A=10$ with an accuracy in
binding energies of 1\% or better;\cite{CARLSON} the practical limit
with current algorithms is probably $A\approx 12$.

\begin{figure}[t]
%\figurebox{22pc}{15pc}{} % to have a box alone
\caption{Energy spectra in nuclei up to $A=10$ \cite{CARLSON}. 
  In each set of levels, calculations with only the two-body
    Argonne $V_{18}$ potential are on the left and those with a 
    semi-phenomenological
    three-body force included are in the middle.  Experimental levels
    are on the right. 
    \label{fig:spectra}}
\epsfxsize=4.0in % will enlarge or reduce the postscript figures based on the xsize
\epsfclipon
\centerline{\rotatebox{270}{\epsfbox{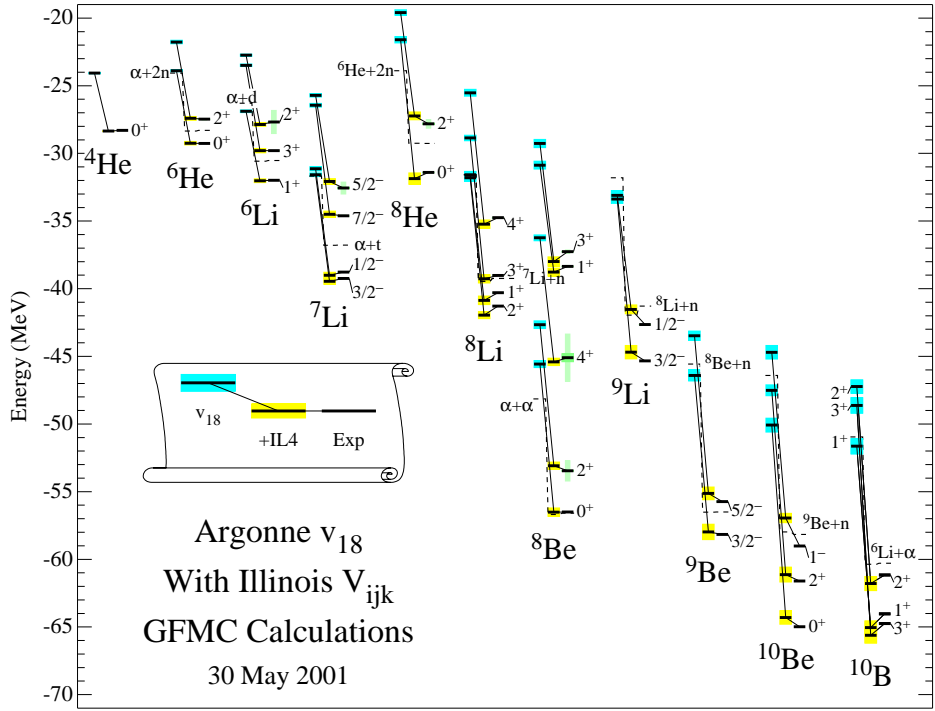}}} % postscript image file name
\end{figure}

The significance of 1\% calculations is that nuclear forces can be unambiguously
tested.  An important result is seen 
by comparing experimental spectra to predictions 
generated using only the two-body potential 
(see Fig.~\ref{fig:spectra}) \cite{CARLSON}.  
The predictions are
underbound and the absolute shortfall grows with $A$, making many-body
forces not only inevitable but quantitatively critical.  Further, a particular
isospin dependence from three-nucleon forces (3NF) is needed 
and 1/2 to 2/3 of the
spin-orbit splittings come from three-body forces.  Currently a
semi-phenomenological form is fit to $A\le 8$ energy levels, but there is a
basic problem:  the number of possible operators for 3NF is much larger than
for NN and it is not practical to examine them all.

Chiral EFT comes to the
rescue with a systematic power-counting scheme that predicts the hierarchy of
many-body forces as well as the hierarchy 
for different three-body forces \cite{VANKOLCK}.
As with two-body forces,
at each order in the EFT expansion 
the power counting scheme specifies the additional three-body diagrams
that should be included.
Coefficients of purely short-range three-body terms must be fit to
three-body scattering or bound state observables.  Note that because
the description is model independent, a fit to one set of observables
can be applied to any other set of observables.
An issue of current study (and controversy)
is whether the $\Delta$ resonance should be
included as a lagrangian degree of freedom \cite{VANKOLCK}.

The next step is to adapt the chiral EFT calculations for GFMC, which
requires local versions of the EFT potential to be constructed.  
Then $A\ge 4$ predictions with a
connection to QCD will be possible!   The future of Monte Carlo for nuclei 
will likely
be Auxiliary Field Diffusion Monte Carlo (AFMDC) with spin-dependent
interactions between nucleons replaced by interactions between nucleons and
auxiliary fields.  The great advantage 
is that one can sample spins and isospins along with
coordinates, so that the cost grows only as $A^3$;  
up to $A\approx 100$ may be
feasible. 

\subsection{The Shell Model Revisited}

\begin{figure}[t]
%\figurebox{22pc}{15pc}{} % to have a box alone
\caption{On the left is 
a schematic representation of full and shell model (SM) spaces.  
The $P$ space is up to $\Lambda_{\rm SM}$ and the $Q$ space
is from $\Lambda_{\rm SM}$ to $\Lambda_{\infty}$.  
On the right is
the magnetic dipole $M1$ form factor for $^3$He calculated in the
full space and with bare operators in various truncated spaces 
 \cite{HAXTON}.  
The curves using effective operators for the same spaces are indistinguishable
from the full (solid) curve.
\label{fig:effops}}
\hbox{\raisebox{1in}{\epsfclipon\epsfxsize=1.4in\epsfbox{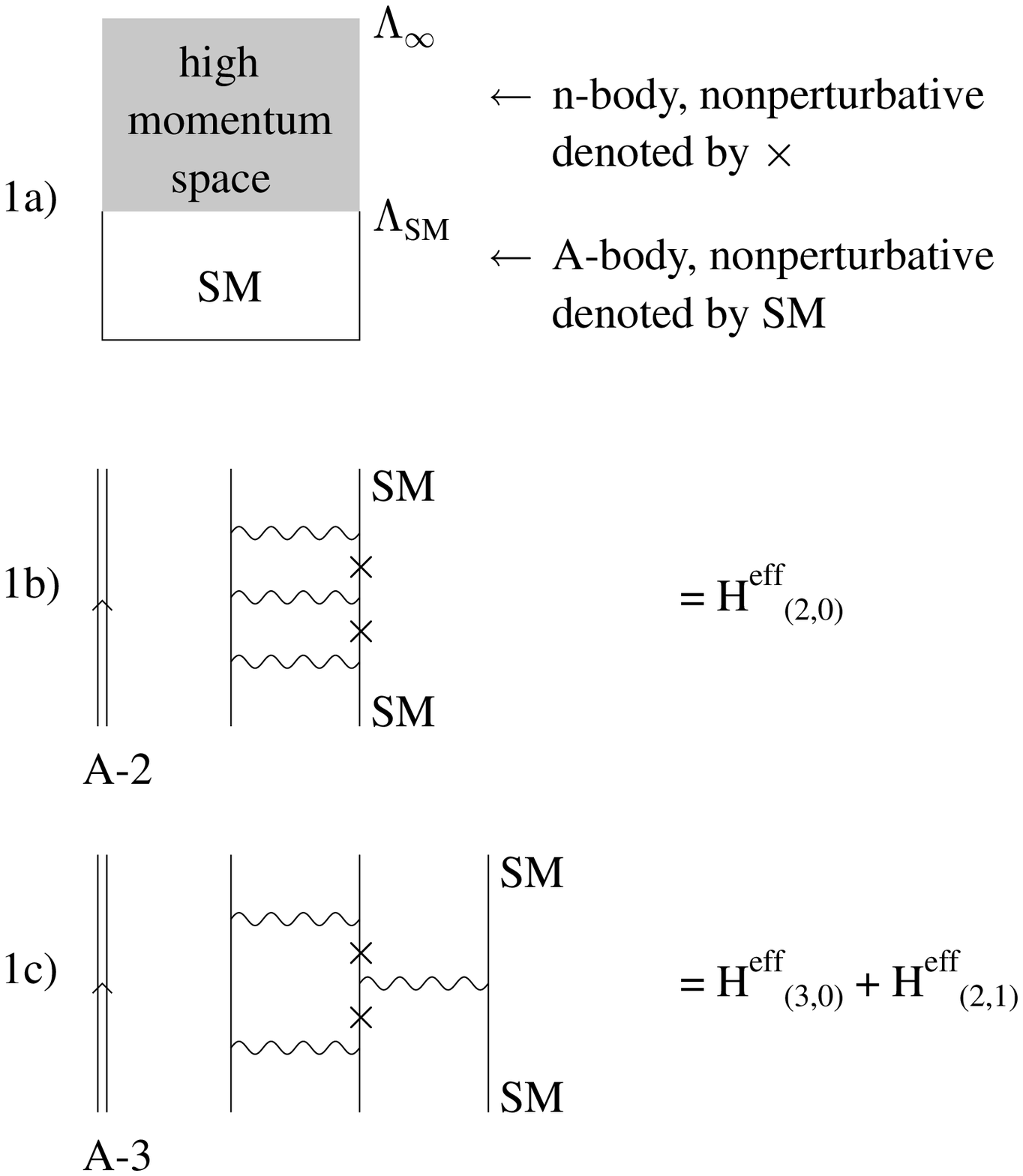}}%
     \hspace*{.1in}\epsfclipon\epsfxsize=3.5in \epsfbox{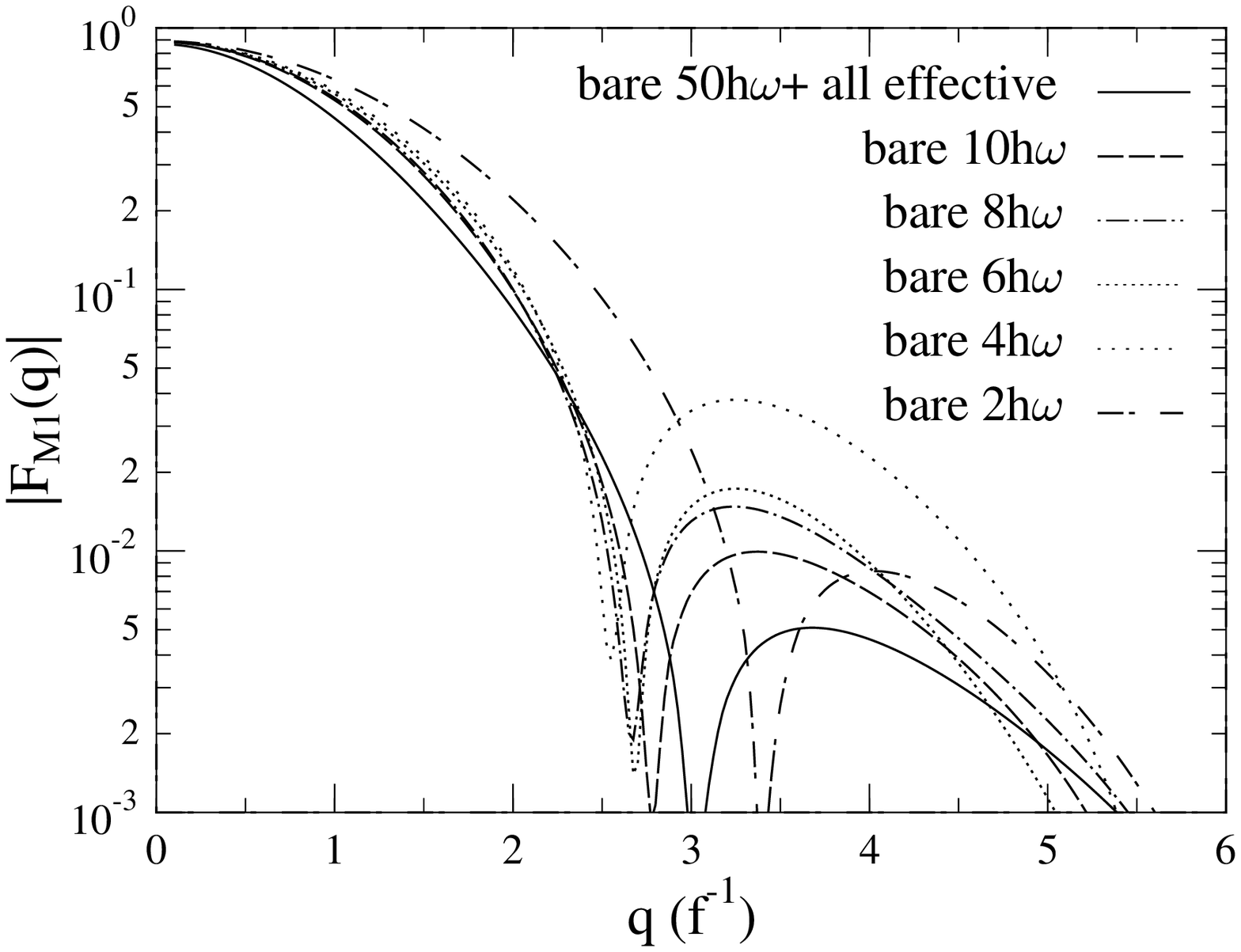}} 
  \vspace*{-20pt}
\end{figure}

In the conventional nuclear shell model, a given free-space Hamiltonian
in the full Hilbert space ($P+Q$) is transformed to an effective Hamiltonian
that is diagonalized in a truncated model
space ($P$), with the same results for low-energy  observables
(see the left side of Fig.~\ref{fig:effops}).  A harmonic oscillator basis is used to
separate relative and center-of-mass (COM) motion 
(recall that there is no fixed COM for nuclei).  A similarity transformation may
be used to decouple the $P$ and $Q$ space or, in practice, a phenomenological
interaction might be used.

There are many problems with actual shell-model implementations, 
which date from an
earlier era in which computational limitations necessitated severe compromises.
Even if only a two-body potential is used in the $P+Q$ space, the
removal of high-momentum states invariably generates many-body forces; in
practice only two-body effective interactions are used.
Another problem 
is sensitive dependence on unphysical parameters, such as the cutoff
$\Lambda_{\rm SM}$ for the
$P$ space or ``starting energies''.
Perhaps most importantly, not only the Hamiltonian but all operators in the 
$P$ space must be effective operators; however, either bare operators are
used or a naive and uncontrolled renormalization (e.g., multiplicative
scaling) is applied.

The shell model approach is a natural setting for renormalization group
methods.  The first steps to take conventional nuclear models to controlled
EFT counterparts have been taken recently by Haxton and 
collaborators \cite{HAXTON}.
They have developed a self-consistent Bloch-Horowitz approach using Lanczos
methods.  The Lanczos construction means that the expensive part of the
calculation need only be performed once and then effective Hamiltonians {\em
and\/} operators for each state can be generated cheaply.
The importance of effective operators is illustrated by the
plot on the right in Fig.~\ref{fig:effops}.
With bare operators, the form factor for different $P$ spaces
is almost random for momentum transfers
above 2\,fm$^{-1}$, while even at low momentum the renormalization
is clearly not a simple scaling, as is often assumed.
In contrast, if effective operators are used, results are independent
of the truncation of the space.

To extend the Bloch-Horowitz approach to heavier nuclei, it is necessary to
effectively decrease the upper bound used to define the
full space, $\Lambda_\infty$.  
One might hope that perturbation theory could be
used to include the effect of higher energy states.  Even the simplest
calculation (e.g., the binding energy of the deuteron) shows that the
nonperturbative effects of the hard-core repulsion in typical NN potentials
spoils any hope of rapid convergence \cite{HAXTON}.  
However, according to the EFT precepts,
the singular contributions from the potential are not resolved in the low-energy
shell-model space, so we can replace them by something simpler:  separable
contact interactions and derivatives.
In practice this leads to an expansion of the harmonic oscillator matrix
elements (Talmi integrals) in powers of the hard core radius over the oscillator
potential \cite{HAXTON}.  

This expansion is matched in the full space and the coefficients 
obtained can then be
evolved to lower $\Lambda$ analytically using a shell renormalization group
equation.  Perturbation theory can be used on the remaining soft potential;
convergence at leading order is rapid \cite{HAXTON}.   
Currently various details of the NLO implementation 
and beyond are being worked out, 
and then the formalism will be embedded in heavier
nuclei.  Eventually the starting potential in the $\Lambda_\infty$ space may be
taken from the chiral EFT, furthering the connection to QCD.

\subsection{Power Counting for Density Functionals}

The large-$A$ region of the table of nuclides requires a different plan of
attack.  The most widely used approaches are referred to as ``self-consistent
mean field'' calculations, but this is a misleading nomenclature, which
implies a Hartree or Hartree-Fock calculation of some underlying interaction.
In fact, it is more appropriate to identify these calculations as Kohn-Sham
density functional theory (DFT), in which the full Hartree plus
exchange-correlation functionals (and not Hartree alone!) are approximated by a
parametrized form.  ``Mean field'' in this context really means a 
limited form of the analytic and nonlocal structure in the energy
functional.

The role of EFT methods here starts with a power counting for Skyrme or
covariant energy functionals.  We consider Skyrme-type functionals here;
the case of covariant functionals is addressed in Brian Serot's talk
\cite{SEROT}.  The functional (given here for $N=Z$ only)
is built from densities ($\rho$), kinetic energy densities ($\tau$),
and currents (${\bf J}$), all obtained as sums over
occupied Kohn-Sham single-particle wave functions: \cite{RING}
\begin{eqnarray}
  \epsilon[\rho,\tau,{\bf J}] 
    &=& {1\over 2M}\tau + {3\over 8} t_0 \rho^2
  + {1\over 16} t_3 \rho^3
 + {1\over 16}(3 t_1 + 5 t_2) \rho \tau  \nonumber
  \\ & & \null
  + {1\over 64} (9t_1 - 5t_2) (\grad \rho)^2  
  - {3\over 4} W_0 \rho \grad\cdot{\bf J}
  + {1\over 32}(t_1-t_2) {\bf J}^2  \ , 
\end{eqnarray}
with
typical [e.g., SkIII] model parameters (in conventional units):
$t_0=-1129$, $t_1=395$, $t_2=-95$, $t_3=14000$, and $W_0=120$  \cite{RING}.
Such large and unsystematic values imply that  
errors from omitted terms may be uncontrolled.

However,
after applying the same Georgi-Manohar NDA that accounts for the chiral EFT
coefficients [Eq.~(\ref{eq:NDA})], we obtain the {\em scaled\/} energy 
functional: \cite{HACKWORTH}
\begin{eqnarray}
   \epsilon[\rho,\tau,{\bf J}] 
 &=&
  {\tau\over \Lambda} \Bigl( c_1 +  c_4 \bar{\rho} + 
	{ c_8 \bar{\rho}^2}\Bigr)
   + {\rho^2\over \fpi^2}\Bigl( c_2 
  + c_3 \bar{\rho} + { c_9 \bar{\rho}^2}\Bigr) \nonumber \\ 
 & & \null +  {(\grad \rho)^2 \over \fpi^2\Lambda^2}  
    \bigl( c_5 + { c_{10} \bar{\rho}}\bigr)	
  +  {\rho \grad\cdot{\bf J}\over \fpi^2\Lambda^2}\bigl( c_6 + 
	{c_{11} \bar{\rho}}\bigr)  
 + \cdots  
\end{eqnarray}
with {\em natural\/} coefficients:
$c_1=0.5$,  $c_2=-0.48$, $c_3=1.05$, $c_4=1.14$,
     $c_5=1.62$, $c_6=-2.31$, and $c_7=0.39$.
The naturalness of the coefficients 
means that truncation errors from
the next order of coefficients can be estimated reliably.  
Additional fits to nuclei have
validated these estimates and the robustness of the power counting.
Work is in progress to embed the Kohn-Sham DFT framework within an
EFT framework using an effective action formalism.

%\vspace*{.1in}

The path for an {\it ab initio\/} 
QCD calculation of heavy nuclei is becoming cleared,
although construction is far from complete.
The key will be to obtain the coefficients
for the Kohn-Sham DFT/EFT not from data but from matching to the chiral
effective field theory.  In turn, those coefficients will not be taken from data
but calculated from matching to nonperturbative lattice QCD calculations.
The latter step may seem far-fetched at this point in time, but recent
work using partially quenched lattice QCD 
to calculate coefficients for pion chiral
perturbation theory is very encouraging \cite{SHARPE}.  A compelling 
feature is that the matching
does not have to occur for a physical experiment or indeed 
even for a physical situation.  
An example is the partially quenched lattice calculation, in which valence
quark masses are different from those in the quark-antiquark sea.

\section{New Ideas for Many-Body Problems}

While renormalization group ideas are widespread in condensed matter many-body
theory, the different 
flavor of EFT approaches used in particle/nuclear theory 
can give new insight into familiar many-body problems.   The dilute fermi gas
(e.g., hard sphere of radius $R$) is a case in point.  What happens if we apply
the EFT program?

We probe the system at low resolution ($k\ll \Lambda \equiv 1/R$), so all
of the physics is short distance and we can use the analog of a multipole
expansion.  Thus our EFT lagrangian is the most
general set of local (contact) interactions: \cite{HAMMER00}
\begin{eqnarray}
   {\cal L}_{\rm eff}   &=& 
       \psi^\dagger \Bigl[i\frac{\partial}{\partial t} 
               + \frac{\nab^{\,2}}{2M}\Bigr]
                 \psi - \frac{{C_0}}{2}(\psi^\dagger \psi)^2
            + \frac{{C_2}}{16}\bigl[ (\psi\psi)^\dagger 
                                  (\psi\!\galnab\!{}^2\psi)+\mbox{ h.c.} 
                             \bigr]   
  \nonumber \\[5pt]
   & & \null +
         \frac{{C_2'}}{8}(\psi\! \galnab\! \psi)^\dagger \cdot
              (\psi\!\galnab\!\psi)
   - \frac{{D_0}}{6}(\psi^\dagger \psi)^3 +  \ldots
  \label{lag}
\end{eqnarray}
There are infinitely many choices for a regularization/renormalization
scheme; which should we choose?

Ideally, only the short-distance scale $\Lambda$ would be in the coefficients
(besides an overall factor of the fermion mass), with no auxiliary
scales.  This would make dimensional analysis a powerful, predictive tool
and optimize our power counting prescription.
However, most regulators such as a cutoff or form factor in
a model potential will introduce a cutoff scale $\Lambda_c$, which can
appear both in the coefficients and the loop integrals.  That allows
arbitrary functions of the dimensionless variables $\Lambda_c/\Lambda$ and
$\kf/\Lambda_c$, which mean that different orders in the $\kf$ expansion
are mixed in any given diagram.
We would like 
a regularization and renormalization prescription that does not
introduce any new scales (except possibly in logarithms).

These conditions are satisfied by dimensional regularization and minimal
subtraction (DR/MS).
A simple matching to the effective range expansion for two-body scattering
determines the two-body coefficients ($C_i$) to any desired order;
three-body and higher coefficients ($D_i$, \ldots) must be determined
by matching to many-body observables.
The consequence of this scheme
at finite density are simple Hugenholtz diagrams for
the energy density at $T=0$, with each diagram contributing to exactly
one order in the $\kf$ expansion \cite{HAMMER00}.
The contribution for each diagram is a coefficient with all of the
$\Lambda$ dependence times a multi-dimensional integral that is simply
a geometric factor (and which is conveniently evaluated even at high
order using Monte Carlo integration).
The absence of an auxiliary scale $\Lambda_c$ means a particularly
clean comparison of particle and hole contributions, which is obscured
in conventional treatments \cite{HAMMER00}.

Another consequence is that the renormalization group equations become a
powerful tool in revealing the analytic structure of observables.
It is well known that the energy density for this system has a term
proportional to $\kf^9 \ln(\kf)$.
The EFT formulation simplifies the identification and
renormalization of such logarithms.
The beta functions are polynomials in the couplings with $\Lambda$-independent
coefficients, which  means that matching powers of $\Lambda$ on 
either side of the renormalization group equations severely restricts
the possible contributions \cite{BRAATEN}.  The bottom line is that one can 
identify the possible powers of logarithms that occur in the
energy density and the corresponding
diagrams that must be inspected for log divergences (see Refs.~\citen{BRAATEN}
and \citen{HAMMER00} for more details).

What if the problem is nonperturbative in the effective field theory
couplings?  (Note that even the dilute hard-sphere example is nonperturbative in
the underlying {\em interaction\/}.)
A dilute Fermi system with a large scattering length $a_s$ provides an
illustrative example.
The EFT power counting described above, when applied to
the large $a_s$ system, implies that {\em all\/}
diagrams with $C_0$ coefficients (two-body contact terms with no derivatives)
must be summed (although all other contributions are perturbative).
At finite density this sum is intractable, so we are led to search for an
additional expansion parameter.
Using DR/MS makes it transparent that we seek a geometric factor as an
expansion parameter, which
implies expanding in the number of space-time dimensions $D$
(actually, the
expansion parameter turns out to be $1/2^{D/2}$).
The expansion is detailed in Ref.~\citen{STEELE00}, although it has
only been carried out to leading order to date.

\section{Summary}

The nuclear many-body problem has now expanded beyond its traditional
scope to become the QCD many-body problem.
Explorations of the QCD phase diagram are accelerating, with new experimental
results from relativistic heavy-ion collisions and new theoretical attacks
on the frontiers.
Effective field theory approaches are playing a role
throughout, including recent progress on the traditional problem
of calculating nuclei under ordinary conditions in terms of potentials
fit to two- and three-body data.
Although many gaps remain to be filled,
a path for
{\it ab initio\/} calculations of heavy nuclei 
starting from QCD with quarks and gluons is now conceivable.
More generally, the application of EFT methods to  many-body
problems promises insight into the analytic structure of observables,
the identification of new expansion parameters, and
a consistent organization of many-body corrections,
with reliable error estimates.

\section*{Acknowledgments}
The author thanks H.-W.\ Hammer, U.~Heinz, R.~Perry, and B.~Serot
for useful comments.
This work was supported in part by the U.S.\ National Science Foundation 
under Grant Nos.~PHY-9800964 and PHY-0098645.

\end{document}